\documentclass[12pt]{iopart}

\usepackage{amssymb}
\begin{document}

\title[]{Scattering of massless waves with arbitrary spin: a unified analysis for Schwarzschild-type medium black holes}

\author{Zhong-Heng Li}

\address{Black Hole and Gravitational Wave Group,
 Zhejiang University of Technology,
 Hangzhou 310023, China}
\ead{zhli@zjut.edu.cn}
\vspace{10pt}

\begin{abstract}
A unified equation is employed to analytically investigate the scattering of massless spin particles by a Schwarzschild-type medium black hole. It is found that for spin particles, curved spacetime induces an effective complex potential analogous to a Coulomb field. While the real part of this potential contributes a real logarithmic term to the phase, the imaginary part gives rise to a corresponding imaginary logarithmic term. Crucially, this imaginary term is precisely responsible for generating the correct asymptotic decay of the wave function. From this framework, a unified analytical expression for the differential cross section is derived, applicable to all particle types considered. Given the successful fabrication of a Schwarzschild-equivalent medium via transformation optics, our theoretical scattering predictions can be tested experimentally by transmitting plane electromagnetic waves through such a structure. Insights gained from these experiments could, in turn, shed light on the scattering of other massless fields (e.g., gravitational waves) by actual black holes.
\end{abstract}

%
%
%
%
%

\section{Introduction}\label{sec:1}
It is well known that scattering studies provide most of our knowledge about particle interactions, from fundamental particles to condensed matter, thereby highlighting the central importance of scattering problems. Research on black hole scattering enhances our understanding of black hole physics and wave propagation in curved spacetimes. Thus, the scattering of waves by black holes is fundamentally significant to both fields.

The study of black hole scattering spans several decades, with scattering by Schwarzschild black holes having been extensively investigated from numerous perspectives [1-16]. In contrast, studies on analogue Schwarzschild black holes remain scarce. To our knowledge, only Ref. [3] has addressed this by substituting the effective refractive index of Schwarzschild spacetime into Maxwell's equations to analyze electromagnetic wave scattering. However, that work neither identified the contribution from the imaginary part of the potential nor extended the analysis to waves of other spin particles. A key challenge in such studies is that the Schwarzschild-type medium metric is typically expressed in isotropic coordinates. In this coordinate system, the dynamical equations for particles become considerably more complex and difficult to solve, which likely explains why scattering by Schwarzschild-equivalent media has rarely been explored.

Over the past decade, breakthroughs in visible-light transformation optics have spurred significant advances in analog gravity [17-20]. Notably, researchers have successfully used photonic chips as an experimental platform to realize optical media whose refractive index distributions mimic the gravitational field of Schwarzschild black holes [17]. Operating at micrometer scales, these photonic chips provide an excellent system for simulating black hole spacetimes, which underscores the importance of studying scattering in Schwarzschild-equivalent media. Accordingly, a primary goal of this paper is twofold: first, to establish a theoretical foundation for electromagnetic wave scattering by black holes in such photonic-chip systems; second, to leverage insights from electromagnetic scattering experiments to infer the scattering characteristics of other massless particles by black holes. To achieve these objectives, we employ a unified theoretical framework to analyze waves of all massless spin particles, thereby establishing a correspondence between electromagnetic waves and their counterparts with other spins.

\section{Schwarzschild-equivalent media}
In isotropic coordinates, the Schwarzschild metric  is given by [21]
\begin{eqnarray}
\mathrm{d}s^{2}=\frac{(1-M/2r)^{2}}{(1+M/2r)^{2}}\mathrm{d}t^{2}
-(1+\frac{M}{2r})^{4}(\mathrm{d}r^{2}+r^{2}\mathrm{d}\theta^{2}+r^{2}\sin^{2}\theta \mathrm{d}\varphi^{2}),
\label{eq:1}
\end{eqnarray}
where $M$ is the mass of the black hole. The event horizon in isotropic coordinates is located at
\begin{equation}
r_{H}=\frac{M}{2}.
\label{eq:2}
\end{equation}
It is well-known that in the standard Schwarzschild metric, the most salient feature at the event horizon is the reversal of the roles of $t$ and $r$ as timelike and spacelike coordinates. However, this argument does not hold when isotropic coordinates are adopted. As Metric (1) shows, the $t$ direction remains timelike and the $r$ direction spacelike both in the region $r>M/2$ and in $r<M/2$. In fact, metric (1) does not cover the interior region of the Schwarzschild black hole; rather, it covers the exterior region twice [22].

If we set
\begin{equation}
\Xi(r)=(1+\frac{M}{2r})^{2},
\label{eq:3}
\end{equation}
and
\begin{equation}
n(r)=(1+\frac{M}{2r})^{3}(1-\frac{M}{2r})^{-1},
\label{eq:4}
\end{equation}
then metric (1) takes the form
\begin{eqnarray}
\mathrm{d}s^{2}=\Xi^{2}(r)[\frac{1}{n^{2}(r)}\mathrm{d}t^{2}-(\mathrm{d}r^{2}+r^{2}\mathrm{d}\theta^{2}+r^{2}\sin^{2}\theta \mathrm{d}\varphi^{2})].
\label{eq:5}
\end{eqnarray}
Equation (5) is the metric describing the Schwarzschild-equivalent medium [23], originally introduced by Gordon [24]. Here, $\Xi$ and $n$ denote the conformal factor and the refractive index, respectively. This metric is important as it has been implemented in an optical medium to construct an analog gravitational system, such that studying electromagnetic wave propagation within it allows for the simulation of spin particles manifesting Maxwell-like wave properties in Schwarzschild spacetime.

\section{ Unified wave equation and spin-weighted spherical harmonics }
We now analyze perturbations of massless fields with spin $s \leq 2$ on the Schwarzschild-equivalent medium background. Metric (1) belongs to type D [25], a class in which the perturbation equations for massless fields of spin $1/2, 1, 3/2$, and $2$ - specifically, the Weyl neutrino, electromagnetic, massless Rarita-Schwinger, and gravitational fields - are fully decoupled [26, 27].

Consequently, each spin state $p$ yields a distinct decoupled equation, resulting in a total of eight equations. Denoting the corresponding wave function as $\Phi_{p}$, our work [28] demonstrates that the equations for spins $0, 1/2, 1, 3/2$, and $2$ in this spacetime can be unified, i.e. through the transformation
\begin{equation}
\Phi_{p}=(\Xi r)^{(p-s)}\Psi_{p}.
\label{eq:6}
\end{equation}
all equations reduce to an elegant form (source-free case) [29]:
\begin{equation}
[(\nabla^{\mu}+pL^{\mu})(\nabla_{\mu}+pL_{\mu})-4p^{2}\psi_{2}+\frac{1}{6}R]\Psi_{p}=0,
\label{eq:7}
\end{equation}
where
\begin{eqnarray}\large
L^{t}&=&\frac{n}{\Xi^{2}}(\frac{n'}{n}+\frac{1}{r}),\nonumber\\
L^{r}&=&\frac{1}{\Xi^{2}}(2\frac{\Xi'}{\Xi}-\frac{n'}{n}-\frac{1}{r}),\nonumber\\
L^{\theta}&=&0,\nonumber\\
L^{\varphi}&=&-\frac{1}{\Xi^{2}r^{2}}\frac{\mathrm{i} \cos\theta}{\sin^{2}\theta};
\label{eq:8}
\end{eqnarray}
\begin{eqnarray}
\psi_{2}=\frac{1}{6\Xi^{2}}[-\frac{n''}{n}+2(\frac{n'}{n})^{2}+\frac{1}{r}\frac{n'}{n}],
\label{eq:9}
\end{eqnarray}
\begin{eqnarray}
&&R=0,
\label{eq:10}
\end{eqnarray}
where the prime denots the derivative with respect to $r$.

Equation (7) reveals that, despite their fundamentally different physical natures, all massless spin particles described by the wave functions $\Psi_{p}$ obey the same dynamical (wave) equation in a Schwarzschild-equivalent medium. This profound similarity in the governing equations is both surprising and remarkable.

As the fundamental result of our analysis, Equation (7) provides the basis for systematically studying the wave-functional similarities among different massless spin particles within this analog medium.

In scattering problems it is customary to consider monochromatic waves with a given frequency $\omega$. Therefore,  the general solution to equation (7) can be written in the form
\begin{eqnarray}
\Psi_{p}=\sum_{l,m}a_{lm}\mathrm{e}^{-\mathrm{i}\omega t} {_{p}R_{lm}}(r) {_{p}Y_{lm}}(\theta,\varphi).
\label{eq:11}
\end{eqnarray}
where $a_{lm}$ is an expansion coefficient and ${_{p}Y_{lm}}(\theta,\varphi)$ denotes a spin-weighted spherical harmonic, with the functional form [30, 31]
\begin{eqnarray}
{_{p}Y_{lm}}(\theta,\varphi)&=&{_{p}S_{lm}}(\theta)\mathrm{e}^{\mathrm{i} m \varphi}
=\big[\frac{(2l+1)}{4\pi}\frac{(l+m)!(l-m)!}{(l+p)!(l-p)!}\big]^{1/2}\mathrm{e}^{\mathrm{i}m\varphi}\big(\sin\frac{\theta}{2}\big)^{2l}\nonumber\\
&\cdot &\sum_{k}\left(\begin{array}{lll}
l-p\\
\,\,\,\,k
\end{array}
\right)\left(\begin{array}{lll}
\,\,\,\,\,\,l+p\\
k+p-m
\end{array}
\right)(-1)^{l-k-p}\big(\cot\frac{\theta}{2}\big)^{2k+p-m},
\label{eq:12}
\end{eqnarray}
where $l$, and $m$ are integers satisfying the inequalities $l\geq s$, $-l\leq m\leq l$.

The spin-weighted spherical harmonics obey key relations including the orthonormality and completeness relation
\begin{equation}
\int_{0}^{2\pi}\mathrm{d}\varphi\int_{0}^{\pi}\mathrm{d}\theta\,{_{p}Y_{lm}}(\theta,\varphi)\,{_{p}\bar{Y}_{l'm'}}(\theta,\varphi)\sin\theta=\delta_{ll'}\delta_{mm'},
\label{eq:13}
\end{equation}
as well as others such as
\begin{equation}
{_{-s}Y_{lm}}(\theta,\varphi)=(-1)^{l+m}{_{s}Y_{lm}}(\pi-\theta,\varphi),
\label{eq:14}
\end{equation}
and
\begin{equation}
\sum_{m=-l}^{l}\mid\,{_{p}Y_{lm}}(\theta,\varphi)\mid^{2}=\frac{2l+1}{4\pi}.
\label{eq:15}
\end{equation}

It is important to note that the angular part of the wave function, ${_{p}Y_{lm}}(\theta,\varphi)$, is universal for all spherically symmetric spacetimes. Only the radial part, ${_{p}R_{lm}}(r)$, is affected by the specific form of the metric functions, and it can be expressed in the following form:
\begin{equation}
{_{p}R_{lm}}(r)=\Xi^{2p-1}n^{(1-2p)/2}r^{-(p+1)}{_{p}Z_{lm}}(r).
\label{eq:16}
\end{equation}
where ${_{p}Z_{lm}}(r)$ is determined by the following equation:
\begin{eqnarray}
\frac{\mathrm{d}^{2}{_{p}Z_{lm}}(r)}{\mathrm{d}r^{2}}&+&\big[\omega^{2}n^{2}+2\mathrm{i}\omega p n(\frac{n'}{n}+\frac{1}{r})-\frac{1}{3}(2p-1)(2p+1)\nonumber\\
&\cdot &\big(\frac{1}{2}\frac{n''}{n}-\frac{1}{4}(\frac{n'}{n})^{2}+\frac{1}{r}\frac{n'}{n}\big)-\frac{l(l+1)}{r^{2}}\big]{_{p}Z_{lm}}(r)=0.
\label{eq:17}
\end{eqnarray}

This equation shows that the functions ${_{p}Z_{lm}}(r)$ is not affected by the conformal factor $\Xi(r)$. Substituting Eq. (4) into Eq. (17), we obtain
\begin{eqnarray}
\frac{\mathrm{d}^{2}{_{p}Z_{lm}}(r)}{\mathrm{d}r^{2}}&+&\big\{\omega^{2}\frac{(1+M/2r)^{6}}{(1-M/2r)^{2}}-\mathrm{i}\omega p\big[\frac{2M(M+2r)^{2}}{r^{2}(M-2r)^{2}}-\frac{(M+2r)^{2}}{2r^{3}}\big]\nonumber\\
&-&(4p^{2}-1)\frac{4M^{2}}{(M^{2}-4r^{2})^{2}}-\frac{l(l+1)}{r^{2}}\big\}{_{p}Z_{lm}}(r)=0.
\label{eq:18}
\end{eqnarray}
Obtaining an exact solution to this equation is highly challenging. However, for scattering problems, what matters is primarily the asymptotic behavior of its solutions, as we will demonstrate.

\section{Asymptotic behavior of wave functions }

Considering the asymptotic behavior of Eq. (18), we find that for large $r$, it reduces to:
\begin{equation}
\frac{\mathrm{d}^{2}{_{p}Z_{lm}}}{\mathrm{d}r^{2}}+[\omega^{2}+\frac{4M\omega^{2}+\mathrm{i}2p\omega}{r}+\frac{15M^{2}\omega^{2}/2-l(l+1)}{r^{2}}+O(r^{-3})]{_{p}Z_{lm}}=0,
\label{eq:19}
\end{equation}

The function satisfying Equation (18) is expected to remain small for $r_{0}<[l(l+1)]^{1/2}/\omega$ [1]. Consequently, the probability in this region can be assumed to be practically zero. This result is of significant physical importance: it implies that a particle in the state $\Phi_{p}$ is largely insensitive to events within a sphere of radius $r_{0}=[l(l+1)]^{1/2}/\omega$. For large $l$, this justifies neglecting the term $15M^{2}\omega^{2}/2$ and the $O(r^{-3})$ term in Eq. (19). Therefore, we only need to solve the following simplified equation:
\begin{equation}
\frac{\mathrm{d}^{2}{_{p}Z_{lm}}}{\mathrm{d}r^{2}}+[\omega^{2}+\frac{4M\omega^{2}+\mathrm{i}2p\omega}{r}-\frac{l(l+1)}{r^{2}}]{_{p}Z_{lm}}=0.
\label{eq:20}
\end{equation}

The solution to Eq. (20) can be written as:
\begin{eqnarray}
{_{p}Z_{lm}}=\mathrm{e}^{-\mathrm{i}\omega r}r^{l+1}F(l+1-p+\mathrm{i}2\omega M, 2l+2; \mathrm{i}2\omega r),
\label{eq:21}
\end{eqnarray}
where $F$ refers to the degenerate hypergeometric function, also known as the Kummer function. There exists another independent solution to Eq. (20), which we discard because it fails to satisfy the physical requirements for scattering.

For asymptotically large $z$, we employ the expansion:
\begin{eqnarray}
F(a, b; z)&\sim \mathrm{e}^{z}z^{a-b}\frac{\Gamma(b)}{\Gamma(a)}\sum^{\infty}_{k=0}\frac{(b-a)_{k}(1-a)_{k}}{k!z^{k}}\nonumber\\
&+\mathrm{e}^{\pm\mathrm{i}\pi a}z^{-a}\frac{\Gamma(b)}{\Gamma(b-a)}\sum^{\infty}_{k=0}\frac{(a)_{k}(a-b+1)_{k}}{k!(-z)^{k}},
\label{eq:22}
\end{eqnarray}
where the sign of the complex phase is fixed by the argument of $z$. For our specific parameter, $z=\mathrm{i}2\omega r$, which necessitates selecting the positive sign.  At large values of $2\omega r$,  using Eq. (22) in Eq. (21), the leading-order behavior of ${_{p}Z_{lm}}$ is given by
\begin{equation}
{_{p}Z_{lm}}\sim (-1)^{l+1}\mathrm{e}^{-\mathrm{i}\omega\rho}+\mathrm{e}^{2\mathrm{i}\delta_{l}}\mathrm{e}^{\mathrm{i}\omega\rho},
\label{eq:23}
\end{equation}
where
\begin{equation}
\mathrm{e}^{2\mathrm{i}\delta_{l}}=\frac{\Gamma(l+1+p-\mathrm{i}2\omega M)}{\Gamma(l+1-p+\mathrm{i}2\omega M)},
\label{eq:24}
\end{equation}
and
\begin{equation}
\rho=r_{*}+\mathrm{i}\frac{p}{\omega}\ln(2\omega r).
\label{eq:25}
\end{equation}
Here $r_{*}$ is called the tortoise coordinate. It is determined by the equation:
\begin{equation}
\partial^{\mu}v \partial_{\mu}v=0, \quad \partial^{\mu}u \partial_{\mu}u=0,
\label{eq:26}
\end{equation}
where $v$ and $u$ are the Eddington-Finkelstein null coordinates, which take the form
\begin{equation}
v=t+r_{*}, u=t-r_{*}.
\label{eq:27}
\end{equation}

The exact tortoise coordinate for Schwarzschild-equivalent media, derived by substituting Eq. (27) into Eq. (26) with metric (1), is
\begin{equation}
r_{*}=r+\frac{M^{2}}{4r}-2M\ln(2\omega r)+4M\ln[2\omega(r-r_{H})].
\label{eq:28}
\end{equation}
When $r\rightarrow \infty$, we have
\begin{equation}
r_{*}=r+2M\ln(2\omega r).
\label{eq:29}
\end{equation}

From Eqs. (6), (11), (16), and (23), the asymptotic form of the wave function for a massless particle with spin state $p$ at spatial infinity is
\begin{equation}
\Phi_{p}\approx\frac{\mathrm{e}^{-\mathrm{i}\omega t}}{r^{s}}\sum_{l,m}\frac{a_{lm}}{r}[(-1)^{l+1}\mathrm{e}^{-\mathrm{i}\omega\rho}
+\mathrm{e}^{2\mathrm{i}\delta_{l}}\mathrm{e}^{\mathrm{i}\omega\rho}]{_{p}Y_{lm}}(\theta,\varphi).
\label{eq:30}
\end{equation}

\section{Asymptotic expansion of plane waves}

In quantum scattering theory, the factors $\mathrm{e}^{-\mathrm{i}\omega r}$ and $\mathrm{e}^{\mathrm{i}\omega r}$ conventionally represent ingoing and outgoing waves, respectively. However, this simple asymptotic form breaks down in the presence of a Coulomb-type potential. In such cases, the asymptotic phase acquires an essential logarithmic correction [32]. A key insight from black hole scattering theory is that the ordinary radial coordinate $r$ must be replaced by a tortoise coordinate $r_{*}$, which inherently incorporates this logarithmic term [33, 34].

For massless waves with spin in a Schwarzschild-equivalent medium, the effective potential derived from Eq. (20) possesses both real and imaginary components. Crucially, both components exhibit a long-range behavior analogous to the Coulomb potential. Consequently, the asymptotic phase becomes complex: the real part of the potential contributes a real logarithmic term, while the imaginary part contributes an imaginary logarithmic term. Therefore, in Eq. (30), the coordinate $\rho$ (which incorporates this complex logarithmic correction) must be used. Thus, the terms $\mathrm{e}^{-\mathrm{i}\omega \rho}$ and $\mathrm{e}^{\mathrm{i}\omega \rho}$ properly describe the ingoing and outgoing asymptotic waves in this context.

Therefore, the expression for a plane wave propagating in a direction that makes an angle $\gamma$ with the z-axis is
\begin{equation}
\Phi_{plane}\sim \mathrm{e}^{\mathrm{i}\omega\rho(\sin\gamma \sin\theta \sin\varphi+\cos\gamma \cos\theta)-\mathrm{i}\omega t}.
\label{eq:31}
\end{equation}
With $p=0$ (implying $\rho=r_{*}$), Eq. (31) reproduces the result from Refs. [33, 34].

Owing to the orthonormality and completeness of the spin-weighted spherical harmonics [Eq. (13)], the plane wave given in Eq. (31) admits an expansion in terms of these harmonics in the Schwarzschild-equivalent medium.

In the asymptotic limit $r\rightarrow\infty$, this plane wave must asymptotically match the form of Eq. (30), but with constant (ingoing and outgoing) coefficients that differ from those in Eq. (30). Consequently, its spin-weighted spherical harmonics expansion is given by:
\begin{equation}
\mathrm{e}^{\mathrm{i}\omega\rho(\sin\gamma \sin\theta \sin\varphi+\cos\gamma \cos\theta)}=\sum_{l,m}c_{lm}u_{lm}(r\rightarrow\infty)\,{_{p}Y_{lm}}(\theta,\varphi),
\label{eq:32}
\end{equation}
where $c_{lm}$ is a constant and $u_{lm}(r\rightarrow\infty)$ represents the asymptotic form of the radial function.

To compute $c_{lm}$, we treat $\theta$ and $\varphi$ as variables, while $r$ is regarded as a fixed parameter. Using Eq. (13), the ``coefficient'' $c_{lm}u_{lm}(r\rightarrow\infty)$ is given by:
\begin{eqnarray}
c_{lm}u_{lm}(r\rightarrow\infty)&=\int^{\pi}_{0}\mathrm{d}\theta\sin\theta {_{p}S_{lm}}(\theta) \mathrm{e}^{\mathrm{i}\omega\rho\cos\gamma \cos\theta}\nonumber\\
&\cdot\int^{2\pi}_{0}\mathrm{d}\varphi \mathrm{e}^{\mathrm{i}\omega\rho\sin\gamma \sin\theta \sin\varphi-\mathrm{i}m\varphi}\nonumber\\
&=2\pi\int^{\pi}_{0}\mathrm{d}\theta\sin\theta {_{p}S_{lm}}(\theta) \mathrm{e}^{\mathrm{i}\omega\rho\cos\gamma \cos\theta}J_{m}(\omega\rho\sin\gamma\sin\theta),
\label{eq:33}
\end{eqnarray}
where $J_{m}$ is the Bessel function. Using the large-argument asymptotic approximation of $J_{m}$ [35], Eq. (33) becomes:
\begin{eqnarray}
c_{lm}u_{lm}(r\rightarrow\infty)&\approx(\frac{2\pi}{\omega\rho\sin\gamma})^{\frac{1}{2}}\int^{\pi}_{0}\mathrm{d}\theta\sqrt{\sin\theta} {_{p}S_{lm}}(\theta) \nonumber\\ &\cdot[\mathrm{e}^{-\frac{\mathrm{i}}{2}(m+\frac{1}{2})\pi}\mathrm{e}^{\mathrm{i}\omega\rho\cos(\theta-\gamma)}
+\mathrm{e}^{\frac{\mathrm{i}}{2}(m+\frac{1}{2})\pi}\mathrm{e}^{\mathrm{i}\omega\rho\cos(\theta+\gamma)}].
\label{eq:34}
\end{eqnarray}

The integral in Eq. (34) contains a rapidly oscillating factor $\mathrm{e}^{\mathrm{i}\omega r_{*}\cos(\theta-\gamma)}$ (or $\mathrm{e}^{\mathrm{i}\omega r_{*}\cos(\theta+\gamma)}$) for large $\omega r_{*}$. However, near the stationary phase point, these oscillations are suppressed. Therefore, the integral is dominated by the contribution from the vicinity of this point. Applying the stationary phase approximation then yields:
\begin{eqnarray}
c_{lm}u_{lm}(r\rightarrow\infty)&\approx(\mathrm{i})^{m+1}(-1)^{l+1}\frac{2\pi}{\omega r}\,{_{p}S_{lm}}(\pi-\gamma)\nonumber\\
&\cdot[(-1)^{l+1}\mathrm{e}^{-\mathrm{i}\omega\rho}
+(-1)^{l+m}\frac{{_{p}S_{lm}}(\gamma)}{{_{p}S_{lm}}(\pi-\gamma)}\mathrm{e}^{\mathrm{i}\omega\rho}].
\label{eq:35}
\end{eqnarray}

Thus, the asymptotic expression for the plane wave that matches the form of Eq. (30) is:
\begin{eqnarray}
\Phi_{plane}&\approx\frac{\mathrm{e}^{-\mathrm{i}\omega t}}{r^{s}}\sum_{l,m}(\mathrm{i})^{m+1}(-1)^{l+1}\frac{2\pi}{\omega r}\,{_{p}S_{lm}}(\pi-\gamma)\nonumber\\
&\cdot[(-1)^{l+1}\mathrm{e}^{-\mathrm{i}\omega\rho}+(-1)^{l+m}\frac{{_{p}S_{lm}}(\gamma)}{{_{p}S_{lm}}(\pi-\gamma)}\mathrm{e}^{\mathrm{i}\omega\rho}]{_{p}Y_{lm}}(\theta,\varphi).
\label{eq:36}
\end{eqnarray}

\section{ Scattering amplitude and cross section}

Here, the scattering problem is addressed using the partial wave method. This approach enables the scattering amplitude and cross section to be written in terms of the phase shifts $\delta_{l}$.

From a comparison between the first terms of Eqs. (30) and (36), we obtain the expression for the expansion coefficient $a_{lm}$:
\begin{equation}
a_{lm}=(\mathrm{i})^{m+1}(-1)^{l+1}\frac{2\pi}{\omega}\,{_{p}S_{lm}}(\pi-\gamma).
\label{eq:37}
\end{equation}
Using Eqs. (32), (35), and (37), the asymptotic form (30) of the wave function for a massless particle with spin state $p$ at spatial infinity can be expressed as:
\begin{equation}
\Phi_{p}\approx\frac{\mathrm{e}^{-\mathrm{i}\omega t}}{r^{s}}[\mathrm{e}^{\mathrm{i}\omega\rho(\sin\gamma \sin\theta \sin\varphi+\cos\gamma \cos\theta)}
+f(\theta,\varphi)\frac{\mathrm{e}^{\mathrm{i}\omega\rho}}{r}],
\label{eq:38}
\end{equation}
where $f(\theta,\varphi)$ denotes the scattering amplitude, which is
\begin{eqnarray}
f(\theta,\varphi)&=\frac{2\pi}{\omega}\sum_{l,m}(\mathrm{i})^{m+1}(-1)^{l+1}{_{p}S_{lm}}(\pi-\gamma){_{p}Y_{lm}}(\theta,\varphi)\nonumber\\
&\cdot[e^{2i\delta_{l}}-(-1)^{l+m}\frac{{_{p}S_{lm}}(\gamma)}{{_{p}S_{lm}}(\pi-\gamma)}]\nonumber\\
&=\frac{2\pi}{\omega}\sum_{l,m}(\mathrm{i})^{m+1}(-1)^{l+1}{_{p}S_{lm}}(\pi-\gamma){_{p}Y_{lm}}(\theta,\varphi)\nonumber\\
&\cdot[\frac{\Gamma(l+1+p-\mathrm{i}2\omega M)}{\Gamma(l+1-p+\mathrm{i}2\omega M)}-(-1)^{l+m}\frac{{_{p}S_{lm}}(\gamma)}{{_{p}S_{lm}}(\pi-\gamma)}].
\label{eq:39}
\end{eqnarray}
If we set $\gamma=0$, then the scattering amplitude reduces to that of an incident wave traveling in the $z$-direction.

The differential cross section, the most important observable in a scattering problem, measures the visibility of the scattering target from a given angle. It follows directly from the scattering amplitude:
\begin{equation}
\frac{\mathrm{d}\sigma}{\mathrm{d}\Omega}=\mid f(\theta,\varphi)\mid^{2}.
\label{eq:40}
\end{equation}
As the primary observable in scattering experiments, the differential cross section--a quantity independent of beam intensity--characterizes the fundamental interaction behavior.

In the case of a scalar field with $p=0$, we have ${_{0}Y_{lm}}(\theta,\varphi)=Y_{lm}(\theta,\varphi)$, where $Y_{lm}(\theta,\varphi)$ denotes the familiar spherical harmonic. Substituting this into relation (14), which becomes $Y_{lm}(\theta,\varphi)=(-1)^{l+m}Y_{lm}(\pi-\theta,\varphi)$, allows the scattering amplitude to be simplified considerably, leading to the result:
\begin{eqnarray}
f(\theta,\varphi)=\frac{2\pi}{\omega}\sum_{l,m}(\mathrm{-i})^{m+1}Y_{lm}(\gamma,0)Y_{lm}(\theta,\varphi)(e^{2i\delta_{l}}-1).
\label{eq:41}
\end{eqnarray}

The total cross section for scalar field scattering is given by
\begin{equation}
\sigma=\int\mid f(\theta,\varphi)\mid^{2}\mathrm{d}\Omega=\frac{4\pi}{\omega^{2}}\sum_{l}(2l+1)\sin^{2}\delta_{l}.
\label{eq:42}
\end{equation}
The derivation of Equation (42) employs Equation (15). The structure of Equation (42) coincides with that of the total scattering cross-section for a quantum central field, but the physical content of the phase shift $\delta_{l}$ is fundamentally different.

\section{ Discussion and conclusion}

The scattering of all massless spin particles by a black hole realized as a Schwarzschild-equivalent medium is studied using a unified wave equation [28]. This yields a single formula for the scattering amplitudes [Eq. (39)], which in turn produces differential cross sections of identical structure across particle types. Given the successful fabrication of such media via transformation optics, these theoretical predictions are directly testable by transmitting plane electromagnetic waves through the spacetime-equivalent medium. Insights gained from such experiments can further inform the scattering behavior of other massless particles by black holes. For scalar particles specifically, the total cross section takes the same mathematical form as in quantum central-field scattering, despite the distinct physical content of the phase shifts. In the partial-wave expansion, each term corresponds to the cross section for a given angular momentum quantum number $l$. Scattering in a particular partial wave is suppressed when $\delta_{l}=n\pi$, whereas resonance (maximum scattering) occurs at $\delta_{l}=\pi/2$.

The influence of the Schwarzschild-equivalent medium on massless spin particles is described by a complex, Coulomb-like potential (Eq. 20). As indicated by Eqs. (25), (29), and (30), its real part yields a real logarithmic phase shift, while the imaginary part gives an imaginary logarithmic term. Consequently, the waves acquire real logarithmic phase distortions at infinity, and their asymptotic amplitude decay is modified by the imaginary term. Table 1 details how this imaginary term leads to distinct decay behaviors: across different particle types, among different spin states of the same type, and between the incident and outgoing waves of a given spin state. Thus, the complex potential provides a unified explanation for both the wavefront distortion and the varied decay patterns observed.

\begin{table}[htbp]
\centering
\caption{Table 1. Asymptotic behavior of the wave functions given by Eq. (30).\\
Note: The results agree exactly with those presented in Table 2.2 of Ref. [33].
\label{tab:1}}
\begin{tabular}{l|r|c}
\hline
Field quantities & Ingoing waves & Outgoing waves \\
\hline
$\Phi_{0}$ & $Y_{lm}\mathrm{e}^{-\mathrm{i}\omega r_{*}}/r$  & $Y_{lm}\mathrm{e}^{\mathrm{i}\omega r_{*}}/r$ \\
$\Phi_{1/2}$ &${_{1/2}Y_{lm}}\mathrm{e}^{-\mathrm{i}\omega r_{*}}/r$   & ${_{1/2}Y_{lm}} \mathrm{e}^{\mathrm{i}\omega r_{*}}/r^{2}$ \\
$\Phi_{-1/2}$ & ${_{-1/2}Y_{lm}}\mathrm{e}^{-\mathrm{i}\omega r_{*}}/r^{2}$ & ${_{-1/2}Y_{lm}}\mathrm{e}^{\mathrm{i}\omega r_{*}}/r$ \\
$\Phi_{1}$ & ${_{1}Y_{lm}}\mathrm{e}^{-\mathrm{i}\omega r_{*}}/r$ & ${_{1}Y_{lm}}\mathrm{e}^{\mathrm{i}\omega r_{*}}/r^{3}$ \\
$\Phi_{-1}$ & ${_{-1}Y_{lm}}\mathrm{e}^{-\mathrm{i}\omega r_{*}}/r^{3}$ & ${_{-1}Y_{lm}}\mathrm{e}^{\mathrm{i}\omega r_{*}}/r$ \\
$\Phi_{3/2}$ & ${_{3/2}Y_{lm}}\mathrm{e}^{-\mathrm{i}\omega r_{*}}/r$ & ${_{3/2}Y_{lm}}\mathrm{e}^{\mathrm{i}\omega r_{*}}/r^{4}$ \\
$\Phi_{-3/2}$ & ${_{-3/2}Y_{lm}}\mathrm{e}^{-\mathrm{i}\omega r_{*}}/r^{4}$ & ${_{-3/2}Y_{lm}}\mathrm{e}^{\mathrm{i}\omega r_{*}}/r$  \\
$\Phi_{2}$ & ${_{2}Y_{lm}} \mathrm{e}^{-\mathrm{i}\omega r_{*}}/r$  & ${_{2}Y_{lm}}\mathrm{e}^{\mathrm{i}\omega r_{*}}/r^{5}$ \\
$\Phi_{-2}$  & ${_{-2}Y_{lm}}\mathrm{e}^{-\mathrm{i}\omega r_{*}}/r^{5}$  & ${_{-2}Y_{lm}}\mathrm{e}^{\mathrm{i}\omega r_{*}}/r$ \\
\hline
\end{tabular}
\end{table}

\ack
This work was supported by the National Natural Science Foundation
of China under Grant No. 12175198.

\end{document}